\documentclass[aps,prd,twocolumn,superscriptaddress]{revtex4-1}
\usepackage{verbatim}
\usepackage{graphicx}
\usepackage{subfigure}
\usepackage{xcolor}
\begin{document}

%\preprint{}

\title{Enhancing the sensitivity of the LUX-ZEPLIN (LZ) dark matter experiment \\ to low energy signals}


%\author{}
%\thanks{}
%\affiliation{}


%\documentclass[superscriptaddress,altaffilletter]{revtex4-1}
%\begin{document}
%\title{LZ Author List for Paper (Draft 210112T1713(GMT))}
%\date{January 12, 2021 17:13 (GMT)}

% 1 
\author{D.S.~Akerib}
\affiliation{SLAC National Accelerator Laboratory, Menlo Park, CA 94025-7015, USA}
\affiliation{Kavli Institute for Particle Astrophysics and Cosmology, Stanford University, Stanford, CA  94305-4085 USA}

% 2 
\author{A.K.~Al Musalhi}
\affiliation{University of Oxford, Department of Physics, Oxford OX1 3RH, UK}

% 3 
\author{S.K.~Alsum}
\affiliation{University of Wisconsin-Madison, Department of Physics, Madison, WI 53706-1390, USA}

% 4 
\author{C.S.~Amarasinghe}
\affiliation{University of Michigan, Randall Laboratory of Physics, Ann Arbor, MI 48109-1040, USA}

% 5 
\author{A.~Ames}
\affiliation{SLAC National Accelerator Laboratory, Menlo Park, CA 94025-7015, USA}
\affiliation{Kavli Institute for Particle Astrophysics and Cosmology, Stanford University, Stanford, CA  94305-4085 USA}

% 6 
\author{T.J.~Anderson}
\affiliation{SLAC National Accelerator Laboratory, Menlo Park, CA 94025-7015, USA}
\affiliation{Kavli Institute for Particle Astrophysics and Cosmology, Stanford University, Stanford, CA  94305-4085 USA}

% 7 
\author{N.~Angelides}
\affiliation{University College London (UCL), Department of Physics and Astronomy, London WC1E 6BT, UK}

% 8 
\author{H.M.~Ara\'{u}jo}
\affiliation{Imperial College London, Physics Department, Blackett Laboratory, London SW7 2AZ, UK}

% 9 
\author{J.E.~Armstrong}
\affiliation{University of Maryland, Department of Physics, College Park, MD 20742-4111, USA}

% 10 
\author{M.~Arthurs}
\affiliation{University of Michigan, Randall Laboratory of Physics, Ann Arbor, MI 48109-1040, USA}

% 11 
% ...

%\collaboration{The LUX-ZEPLIN (LZ) Collaboration}
%\maketitle
%\end{document}

\date{\today}

\begin{abstract}
Two-phase xenon detectors, such as that at the core of the forthcoming LZ dark matter experiment, use photomultiplier tubes to sense the primary (S1) and secondary (S2) scintillation signals resulting from particle interactions in their liquid xenon target. This paper describes a simulation study exploring two techniques to lower the energy threshold of LZ to gain sensitivity to low-mass dark matter and astrophysical neutrinos, which will be applicable to other liquid xenon detectors. The energy threshold is determined by the number of detected S1 photons; typically, these must be recorded in three or more photomultiplier channels to avoid dark count coincidences that mimic real signals. To lower this threshold: a) we take advantage of the double photoelectron emission effect, whereby a single vacuum ultraviolet photon has a $\sim20\%$ probability of ejecting two photoelectrons from a photomultiplier tube photocathode; and b) we drop the requirement of an S1 signal altogether, and use only the ionization signal, which can be detected more efficiently. For both techniques we develop signal and background models for the nominal exposure, and explore accompanying systematic effects, including the dependence on the free electron lifetime in the liquid xenon. When incorporating double photoelectron signals, we predict a factor of $\sim 4$ sensitivity improvement to the dark matter-nucleon scattering cross-section at $2.5$~GeV/c$^2$, and a factor of $\sim1.6$ increase in the solar $^8$B neutrino detection rate. Dropping the S1 requirement may allow sensitivity gains of two orders of magnitude in both cases. Finally, we apply these techniques to even lower masses by taking into account the atomic Migdal effect; this could lower the dark matter particle mass threshold to $80$~MeV/c$^2$.
\end{abstract}

\pacs{}
%\keywords{}

\maketitle

%\subsection{}
%\subsubsection{}

%%%%%%%%%%%%%%%%%%%%%%%%%%%%%%%%%%%%%%%%%%%%%%%%%%%%%%%%%%%%%%%%%%%%%%%%
\section{Introduction\label{sec:intro}}

The LZ detector is a liquid xenon time projection chamber (LXe-TPC) aiming at the discovery of particle dark matter scattering via the detection of low energy nuclear recoils in its liquid xenon (LXe) target. LZ is being installed at the Sanford Underground Research Facility (SURF), with operation planned to start in 2021 in the Davis Campus (4,300 m.w.e.). Details of the experiment can be found in Refs.~\cite{LZNIM,LZTDR}.

LZ relies on the detection of scintillation and ionization signals generated in the active medium to record particle interactions. This is achieved by instrumenting a two-phase (liquid/gas) system containing 7~tonnes of (active) LXe mass with two arrays of photomultiplier tubes (PMTs), with 241 and 253 units installed in bottom and top arrays, respectively. These detect both a prompt scintillation signal from the liquid (S1) and a delayed electroluminescence response (S2) generated by drifting the ionization released at the interaction site to a thin layer of vapor above the liquid. A drift field of 300~V/cm is applied to the target volume between cathode and gate wire grids in order to drift this charge. A much stronger field of $\approx$10~kV/cm applied to the gas phase allows most of those electrons to be emitted from the liquid surface, and to generate electroluminescence photons as they travel up to an anode grid a few mm away. The two signals allow for fiducialization in $x$,$y$ via the pattern of S2 light detected in the top array, and in $z$ via the time difference between the S1 and S2 optical pulses. The nominal dark matter (DM) search will focus on nuclear recoils within a fiducial mass of $5.6$~tonnes, where external backgrounds are suitably self-shielded.

For recoil energies of $\sim\!1$~keV, only a few photons and electrons are produced by the primary recoil track. The relatively modest photon detection efficiency achieved in the liquid phase ($\sim\!0.1$) impairs the detection of S1 signals, while the high detection efficiency for electrons ensures that the sizable S2 pulses are still recorded at these energies. If the S1 pulse cannot be unambiguously tagged, then the event is classed as `S2-only'. Sources of few-electron background signals contribute to the observed rate of S2-only events, especially those originating at the electrode grids---such as radioactive backgrounds where the scintillation is obscured and spurious electron emission from cathodic electrodes---or electron-train pile-up~\cite{LUXSpurious}. In these events it is not possible to measure the ionization drift time and, hence, $z$-fiducialization is in principle lost. In addition, PMT dark counts (DC) and other artifacts can mimic the missing S1 signal and be randomly associated with the lone S2 pulse to form a viable background event topology. For this reason, the nominal DM search analysis will require a viable S1 to register at least 3 individual photons detected in near-coincidence on different channels (i.e.~a 3-fold PMT coincidence requirement) to avoid a significant rate of such random associations.

The 3-fold coincidence level required of the S1 pulse leads to efficient detection for DM masses above $\sim\!5$~GeV/c$^2$. For lower masses only a very small percentage of particle-induced nuclear recoils will produce sufficient primary scintillation to enable reconstruction and identification of the S1 pulse. The Lee-Weinberg limit~\cite{LeeWeinbergLimit} restricts thermal dark matter models (e.g.~WIMPs) to masses above a few GeV/c$^2$ to avoid the overclosure of the universe. However, other well motivated models offer light candidates on the order of 0.1--10~GeV/c$^2$ that evade this lower bound, such as asymmetric or other hidden sector dark matter models~\cite{ADMreview, ZUREK201491}, and it is desirable to explore this parameter space in LZ. Dropping the S1 requirement altogether may be considered, using only the ionization signals in an `S2-only' analysis~\cite{Xe10S2o,Xe1TS2o}, especially if significant backgrounds from the cathode and gate grids can be controlled and/or estimated robustly. Electron diffusion in the LXe introduces a drift-time dependence to the S2 pulse shape which may be used to partially recover depth information---or at least discriminate between grid emissions and ionization signals from the LXe bulk~\cite{sorensendiffusion,Xenon1T}.

An intermediate possibility is to extend the nominal analysis by lowering the S1 coincidence level while preserving a clean sample of S1-S2 pairs. Frequent random coincidences of PMT dark counts are expected when lowering the S1 coincidence requirement to 2-fold (on the order of hundreds in the full LZ exposure), but Xe scintillation has two properties that allow S1 light to be discriminated from this background source. Xenon scintillates mostly in the vacuum ultraviolet (VUV) range, producing short-wavelength photons capable of emitting two `photoelectrons' from the PMT photocathodes~\cite{Faham2015,DPE}---an effect termed double photoelectron emission (DPE)---thus producing a larger response than is typical for thermionic dark counts. In addition to this wavelength-specific effect, LXe scintillation is relatively fast (decay times up to a few  tens of ns) and, although pulse shape discrimination (PSD) has limited power for nuclear recoil identification in this medium at high fields, the timing of the scintillation signal can nonetheless be used to distinguish it from the above sources of small signals. This was indeed the discrimination mechanism pursued in experiments such as ZEPLIN-I~\cite{ZEPLINI} and XMASS~\cite{XMASSI}, and a study of PSD in LUX has explored the potential of this technique at low energies~\cite{LUXPSD}.

By allowing the 3-fold requirement to be lowered or even completely relaxed, searches for nuclear recoil (NR) and electron recoil (ER) interactions can be conducted at very low energies. Despite the high atomic mass of xenon, LXe-TPCs achieve competitive sensitivity in this regime as a result of having realized remarkably low background rates at the core of their active volumes. This may enable new searches for light ($\sim$GeV/c$^2$) DM particle scattering producing detectable NR signals, or even lower mass particles detected via the ER signals generated through the atomic Migdal effect~\cite{MigdalEffect, LUXMigdal}. In addition, a significant enhancement of the sensitivity to coherent elastic neutrino-nucleus scattering (CE$\nu$NS) may be possible, targeting in particular the very soft NR spectra expected from solar $^8$B neutrinos and from supernova neutrinos.

The LUX experiment has demonstrated the use of DPE signals in a recent reanalysis of 2013 data, lowering the S1 threshold to single detected photons~\cite{LUXDPE}. LZ possesses several characteristics in terms of target size, background, and detector performance that could potentially render this technique more powerful. 

Further in that direction, the XENON1T experiment has recently published S2-only analyses of their data which established leading dark matter constraints for NR interactions~\cite{Xe1TS2o} and via the Migdal effect~\cite{Xe1TMigdal}. We expect LZ to be proportionally more competitive (i.e. even after accounting for the larger target mass) as its grid production was executed under strict cleanliness protocols, thus reducing radon plate-out and the subsequent $^{210}$Pb chain backgrounds. Additionally, spurious electron emission has been extensively studied and passivation treatment applied to the gate grid, where the largest fields are expected~\cite{Tomas2018,slacgridspaper}.

In this article we assess the sensitivity improvements expected from decreasing the S1 detection threshold in LZ through Monte Carlo simulations---either by including DPE-enhanced S1 pulses below the nominal 3-fold threshold, or by dropping the requirement of an S1 in the event altogether. In Sec.~\ref{subsec:S1andDPE} we focus on the S1 pulse in a DPE analysis, and develop the required methodology to optimize the S1 selection for events registering exactly two channels in coincidence. In Sec.~\ref{subsec:S2only} we focus instead on the S2 pulse, and study the potential of an `S2-only' analysis using realistic backgrounds and S2 pulse characteristics. In Sec.~\ref{sec:DPEanalysis} we apply the DPE technique in the standard NR channel and also to ER signals in which the recoiling nucleus is not detected directly but a low-energy electron can be generated through the Migdal effect. We proceed analogously in Sec.~\ref{sec:S2oanalysis} with the S2-only analysis. Finally, we compare the two threshold-reduction techniques and discuss their applicability in Sec.~\ref{sec:conclusions}. LZ baseline parameters and exposure (1,000 days and 5.6--tonne fiducial mass) are assumed~\cite{LZTDR,LZsens}, unless stated otherwise.

%%%%%%%%%%%%%%%%%%%%%%%%%%%%%%%%%%%%%%%%%%%%%%%%%%%%%%%%%%%%%%%%%%%%%%%%%%%%%
\section{Techniques to lower the energy threshold\label{sec:techniques}}

The nominal DM sensitivity analysis relies on S1 and S2 pulse areas to produce probability density functions (PDFs) and to carry out profile likelihood ratio (PLR) hypothesis tests. This analysis requires a 3-fold S1 coincidence level and a minimum raw (uncorrected) S2 signal of approximately 5 extracted electrons ($S2r > 420$~photoelectrons)~\cite{LZsens}. In the following two subsections we describe additional experimental observables that can be exploited to lower the detection threshold of LZ, applicable to other two-phase LXe-TPCs. In Sec.~\ref{subsec:S1andDPE}, we motivate lowering the S1 coincidence threshold from 3 to 2-fold and exploiting the DPE effect and LXe scintillation timing to reject random coincidence backgrounds. Similarly, in Sec.~\ref{subsec:S2only} we explore dropping the requirement of an S1 signal altogether, and using only the S2 signal. The detailed application of these two techniques to LZ simulated data is then presented in Sec.~\ref{sec:DPEanalysis} and Sec.~\ref{sec:S2oanalysis}, respectively.

\subsection{Exploiting photon timing and the DPE effect\label{subsec:S1andDPE}}

\begin{figure*}[htp]
  \centering
  \subfigure{\includegraphics[width=0.49\textwidth]{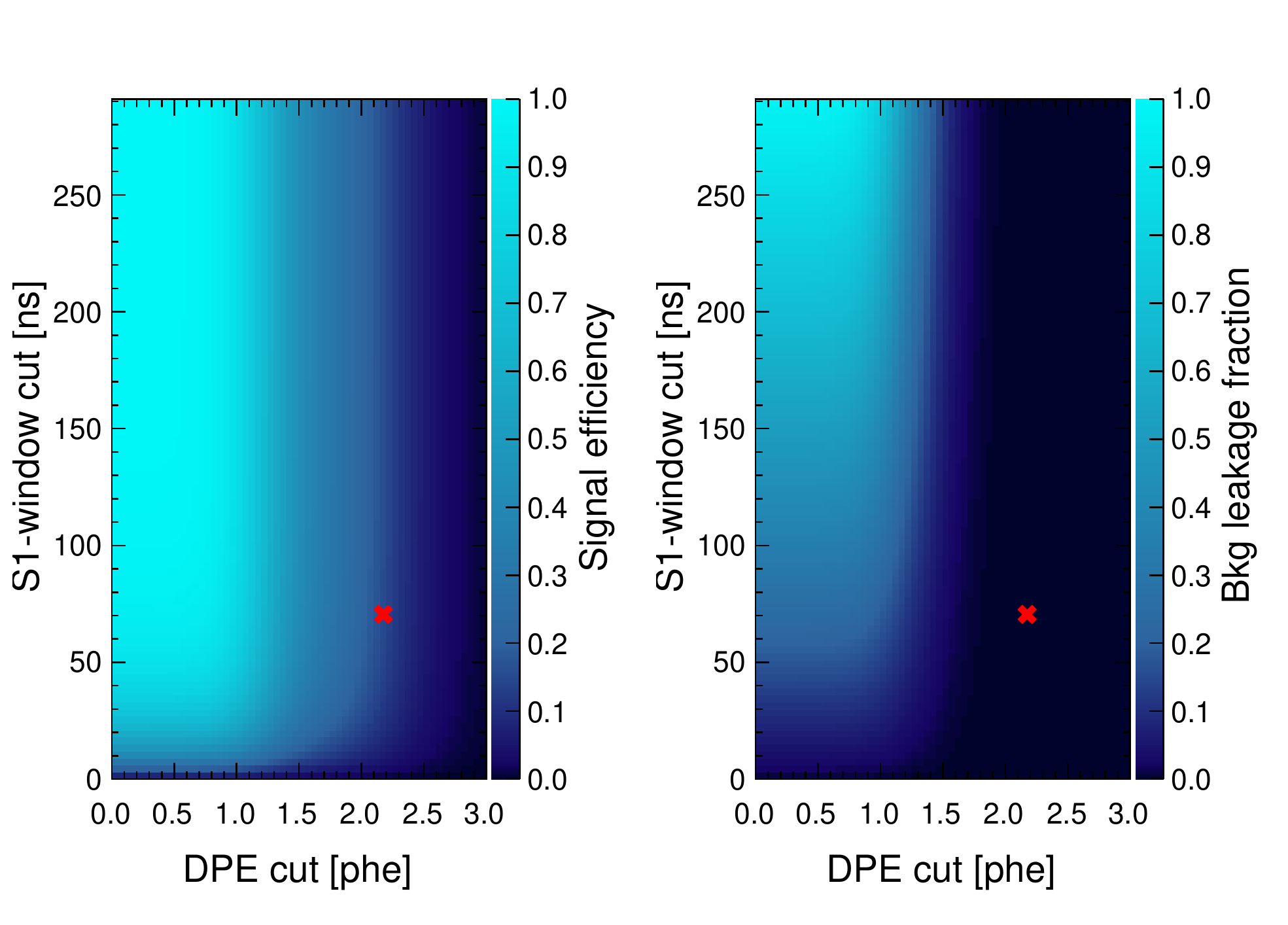}}\quad
  \subfigure{\includegraphics[width=0.49\textwidth]{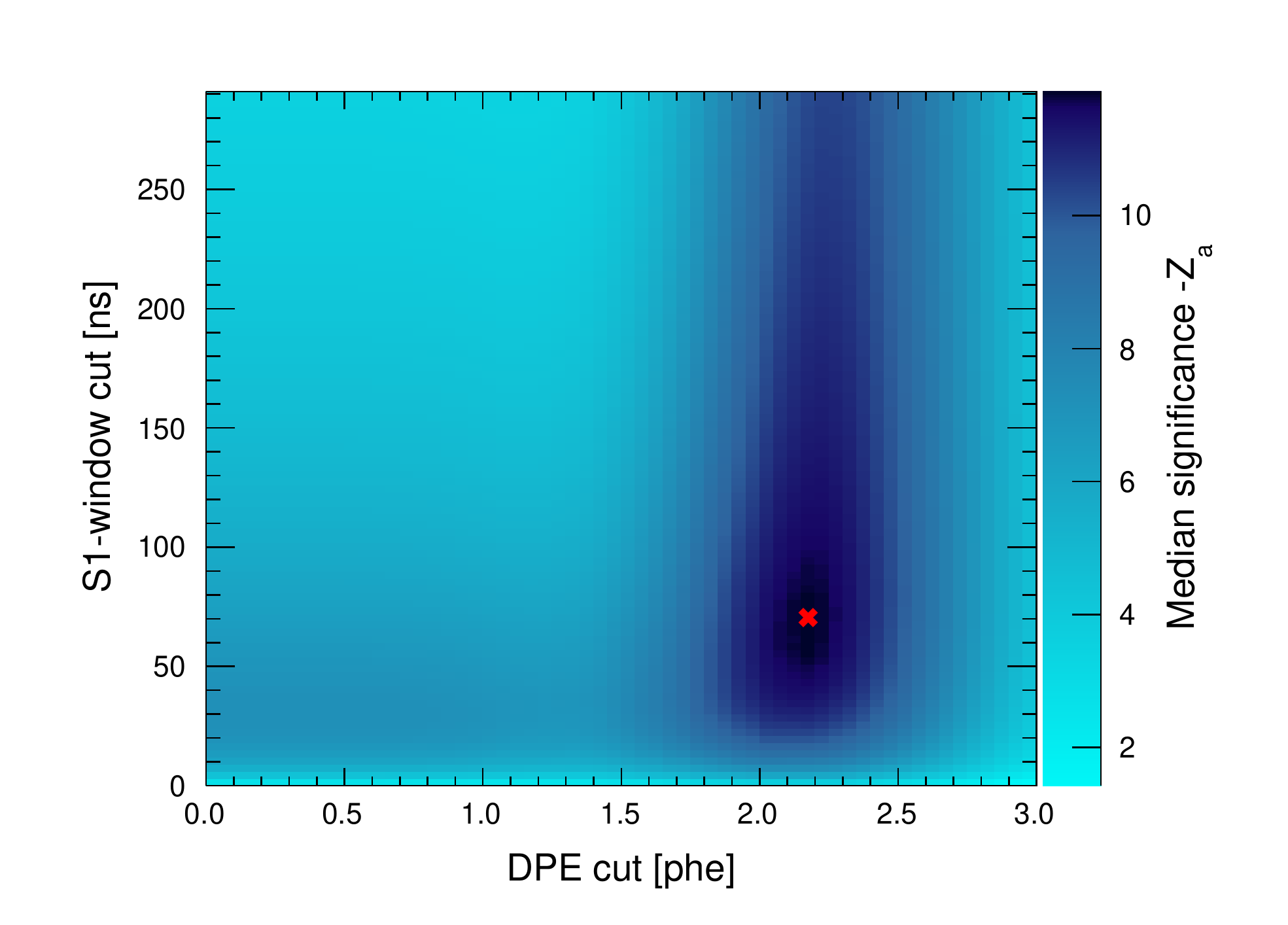}}
  \caption{Two-fold NR signal efficiency ($z$-axis) as a function of the DPE and S1 window cuts (left). Background leakage fraction as a function of those cuts (center). Optimization of the median significance $Z_A$ (Eq.~\ref{eq:asimov}) assuming a minimum NR signal given by ${}^8\textrm{B}$ neutrinos and a background given by random coincidences of PMT dark counts with S2-only events (right). The red cross indicates the result of the optimization.\label{fig:optimisation}}
\end{figure*}

The DPE analysis is based on the fact that we can distinguish real S1 pulses from fake S1-like pulses formed by PMT dark counts due to their different pulse area and timing distributions. We describe here the assumptions behind the production of signal and background PDFs for an analysis requiring a 2-fold coincidence level in S1.

The LXe scintillation spectrum peaks in the VUV region near $175$~nm~\cite{FUJII2015293}. These scintillation photons may interact in a PMT photocathode and give the resulting photoelectron sufficient energy to produce another free electron, which may itself be emitted into the PMT vacuum. If both are successfully amplified, the pulse area---on average twice as large as that obtained from a single photoelectron (SPE)---becomes a clear indication that a VUV photon has been detected. Typically, the area of a pulse is calibrated and expressed in photoelectrons~(phe) as an estimate of the number of photons detected. This calibration is biased in the presence of DPE. We term the unbiased quantity `photons detected'~(phd) and we determine it by photon counting and DPE calibration. The DPE effect has been measured to average $22.6\%$ at LXe temperature for the LZ PMTs (3-inch Hamamatsu R11410-22)~\cite{DPE}.

The LZ dual-gain data acquisition system amplifies PMT waveforms with 60~ns shaping time in its high-gain channels, and digitizes at 100~MS/s (10~ns samples). At nominal PMT gain an SPE detection efficiency in excess of 95\% has been demonstrated~\cite{LZTDR}; in the case of DPE, the probability of detecting this larger pulse is practically 100\%.

Given an S1 with two channels in coincidence, the requirement that at least one exhibits a pulse area consistent with the DPE response can improve discrimination of VUV from other photon-like pulses if the area threshold is optimized (this is hereafter referred to as the `DPE cut'). These photon-like pulses include thermionic dark counts or sources of longer-wavelength light, such as the fluorescence of polytetrafluoroethylene (PTFE), a material which is employed extensively in the detector as a VUV reflector~\cite{LZTDR}. Accidental S1-S2 pairings involving genuine S1-only signals, such as sub-cathode interactions from which ionization cannot be collected, remain an irreducible background, but we expect a negligible contribution at very small S1 signals~\cite{LUXDPE}.

Additionally, the timing distribution of LXe scintillation is distinct from that expected from random coincidences between PMT dark counts (constant rate) or from PTFE fluorescence (which have decay constants which are typically much longer than the LXe scintillation). For very low energy interactions, LXe scintillation is characterized by only two time constants ($\tau_1$ and $\tau_3$), corresponding to the decay of the singlet and triplet states of the Xe$_2^*$ excimer~\cite{LUXPSD}, respectively; these have different coefficients for ER and NR interactions, listed in Table~\ref{tab:S1timing}. An electron-ion recombination component which becomes important for higher-energy electron recoils is negligible in the keV regime at the electric field of interest here~\cite{LUXPSD, LUXCH3T}. Other factors affect the timing of the ER and NR signals which require careful modeling: (a) the dispersion of VUV photon transit times in the detector, which is obtained by Monte Carlo simulation and is approximated by an exponential distribution with $\tau_{\texttt{\tiny OTS}}=1.34$~ns; (b) the PMT transit time spread (TTS), with $\sigma = 3.9$~ns measured for this model~\cite{hamamatsucatalog,Barrow2017}; (c) the timing calibration uncertainty between TPC channels, including dispersion of PMT mean transit times plus cabling, electronics and digitization delays, which will be determined to $\mathcal{O}(\textrm{ns})$ using a dedicated LED calibration system---these are conservatively assumed to be Gaussian with $\sigma = 2$~ns after timing calibrations with the full electronics chain. The LZ experiment will account for these effects via careful timing calibrations alongside the simulation of S1 photon transport and of the electronics response of the detector and its electronics chain.

The background from triple dark count coincidences is negligible for coincidence windows on the order of $100$~ns, and so the choice of S1 window is more flexible for the nominal $\geq\!3$-fold dataset and need not concern us here. Optimization of the S1 coincidence window is important for signals with only two channels in coincidence. Such pulses are unlikely to occur in the same channel, and we define the S1 pulse width for these 2-fold pulses (hereafter `S1 window') as the time between single-photon pulse peak times determined by waveform fitting. We impose a cut on the maximum S1 window to reject random coincidences in favor of scintillation signals.

\begin{table}[b]
\caption{S1 timing parameters published in Ref.~\cite{LUXPSD}. The singlet and triplet ($\tau_{1, 3}$) molecular lifetimes and their relative strengths ($C_1/C_3$) are given for electron and nuclear recoils.}
\label{tab:S1timing}
\begin{tabular}{ c | c | c | c}
\hline
    & $\tau_1$/ns & $\tau_3$/ns & $C_1$/$C_3$ \\ \hline
 ER & $3.27\pm0.66$ & $25.89\pm0.06$ & $0.333\pm0.082$\\  
 NR & $3.27\pm0.66$ & $23.97\pm0.17$ & $1.97\pm0.43$\\   
\hline
\end{tabular}
\end{table}

The time delay distribution for two dark count coincidences is $P(t) = re^{-rt}$~\cite{KNOLL}, where $r$ is the dark count rate, which averages $\sim\!18$~Hz for each R11410 PMT operating in the cold LXe in low-background conditions~\cite{Xe1TDC}. We note that this is an approximate calculation which is likely to include spurious light from sources such as PTFE fluorescence (cf.~discussion in Ref.~\cite{LUXDPE}). These dark count coincidences may occur within the maximum electron drift time ($\sim\!850~\mathrm{\mu}$s) before an S2-only pulse, faking an event with a viable S1-S2 pairing. The frequency of these pairings will be dependent on the S2-only rate, which is a key parameter for both analyses considered here. The LUX and XENON1T experiments have published rates on the order of 1~mHz for such events~\cite{LUXreanalysis,Xe1Tmodels}, and we also assume a $1$~mHz total rate---balancing the larger electrode area against the much stricter grid manufacture program followed in LZ~\cite{Tomas2018, slacgridspaper}. 

As the S1 coincidence level is lowered, so is the energy threshold of the detector. At present, the ionization yield of NR in LXe is known for recoil energies as low as $0.3$~keV~\cite{DDnew}. We use the NEST~\cite{NESTv201} parametrization of LXe yields for all energies, but we impose a model cutoff at $0.3$~keV, and assume NEST predictions for the light yield---based on low energy measurements down to $1.08$~keV~\cite{LUXDD}. Similarly, the ionization and scintillation yields for ER interactions have been calibrated to $0.19$~keV and $1.3$~keV, respectively~\cite{LUXXe127,LUXCH3T}---the value of $0.19$~keV is chosen as the model cutoff and the NEST prediction down to that value is assumed. The relevant experimental yields and the NEST model adopted here (v2.0.1) are depicted in Fig.~6 of Ref.~\cite{LUXDPE}. In the final statistical analysis NEST parameters are varied within their fit uncertainties and correlations to obtain the~$\pm1\sigma$ uncertainties on the charge and light yields. We follow the model in~Ref.~\cite{NESTv201} which does not impose anti-correlation of the two response channels at the lowest energies. A more detailed study of how these systematic uncertainties affect the LZ sensitivity in this regime will be the subject of a forthcoming publication.

The optimization of the S1 window and DPE cuts, depicted in Fig.~\ref{fig:optimisation}, must consider the relative signal-to-background ratio achieved as a function of both parameters. Boron-8 neutrino CE$\nu$NS interactions are virtually indistinguishable from light DM recoils: in the DPE analysis, the signal region fully overlaps with $^8$B for DM particle masses in the range 2--6~GeV/c$^2$ (see Sec.~\ref{sec:DPEanalysis}). For events with an S1 coincidence level of precisely 2 channels, the $^8$B `signal'  predicted using NEST and baseline LZ parameters amounts to $\sim\!100$ counts; we adopt this as the number of signal-like NR events for this optimization. The main background in this particular analysis is due to the random association between a fake S1 pulse created by two coincidental dark counts associating with an S2-only event; this background depends on the corresponding rates and the maximum electron drift time in the TPC. For otherwise nominal parameters and no cuts, this event topology would register 790~counts in the full exposure. Other, more standard NR and ER backgrounds are negligible in comparison. The signal efficiency and background leakage in this optimization are calculated as a function of the S1 window and DPE cuts, with the result shown in Fig.~\ref{fig:optimisation}~(left, center). We optimize based on the median significance calculated with the Asimov test statistic,
\begin{equation}
Z_A = \sqrt{2\left((s+b)\ln\left(1+\frac{s}{b}\right)-s\right)},
\label{eq:asimov}
\end{equation}
where $s$ is the signal count and $b$ is the expected background~\cite{Cowan2011}. This is valid over a wide range of $s$ and $b$, and the function reduces to $s/\sqrt{b}$ for $s \ll b$. Optimal values are found at an S1 window of $70.5$~ns and a DPE cut at $2.17$~phe, as highlighted in Fig.~\ref{fig:optimisation}~(right). These give a signal efficiency of $18.7\%$ and reduce the mean background expectation to only $0.4$~events in the full exposure. These numbers are used in the remainder of the analysis.

This signal efficiency is significant to enable the measurement of steep NR spectra, especially when relying on a small fraction of events for which the S1 signal has benefited from binomial over-fluctuations. Naturally, it is possible to further lower the coincidence requirement to 1-fold. However, to achieve the same background a more aggressive DPE cut is necessary, lowering the signal efficiency to $\mathcal{O}(1\%)$. At this level a significant amount of signal could present triple photoelectron emission (TPE), estimated at $\sim0.6\%$ for the R11410 PMTs~\cite{DPE}. While an order-of-magnitude signal rate increase is possible between 2- and 1-fold, it would be necessary to characterize the TPE probability more precisely to avoid $\mathcal{O}(10\%)$ uncertainty on the signal rate. In LUX this reduction was possible given the lower number of (smaller) PMTs and the much shorter exposure considered in Ref.~\cite{LUXDPE}.

%%%%%%%%%%%%%%%%%%%%%%%%%%%%%%%%%%%%%%%%%%%%%%%%%%%%%%%%%%%%%%%%%%%%%%%%%%%%
\subsection{Exploiting the S2 shape for an S2-only analysis \label{subsec:S2only}}

In an S2-only analysis the pulse size and shape provide the only available information about the interactions, and the assumptions made about these are important in this study. We describe here the assumptions that are made to build PDFs for the S2-only analysis.

Regarding the S2 size, the main parameters to consider are the free electron lifetime in the liquid, the emission probability at the surface, and the photon yield in the gas phase. The LZ gate-anode system was designed to achieve high electroluminescence gain over an $8$~mm gas gap, with the S2 response predicted to yield $83$~phd per electron emitted--- enabling a low analysis threshold at $5$~electrons, a high electron extraction probability at the surface ($\sim90$\%~\cite{Edwards2018,Xu2019}), and the accurate reconstruction of background interactions near the TPC walls~\cite{LZTDR}. Baseline LZ parameters also include an electron lifetime of $850~\mathrm{\mu}$s, corresponding to only one drift length in the TPC, which is rather conservative for such a large detector. This places the ionization threshold at nearly $14$~electrons for the deepest interactions. We will also examine a scenario with twice the electron lifetime ($1.7$~ms), and discuss the impact of this parameter and its measurement uncertainty on both sensitivity calculations.

As stated previously, we expect an S2-only rate of $\mathcal{O}(1~\mathrm{mHz})$. We consider also the impact of a higher rate of $10$~mHz in our analysis. These additional events may be due to radioactivity from the grids and other detector materials near the top and bottom of the active volume, and from spurious emission of multi-electron clusters from gate and cathode~\cite{AdamThesis,LUXSpurious,Tomas2018}. In our treatment we allocate equal emission rates and spectra to the two grids for simplicity: although the gate has higher fields at the wire surfaces and more field-lines connecting to the extraction region, the cathode was not passivated to mitigate electron emission (although the cleanliness procedures were strict also in this case). We describe in Sec.~\ref{sec:S2oanalysis} how the grid-related S2 spectra were estimated for the background model. It is of paramount importance to distinguish these events from those originating in the bulk of the active volume.

Ordinarily, LXe-TPCs rely on the time delay between S1 and S2 to measure the interaction depth and hence fiducialize in the vertical direction. In the absence of S1 information, one may exploit the longitudinal diffusion of drifting electrons in the liquid phase~\cite{sorensendiffusion}, which smears the shape of the S2 pulse in a manner correlated with the interaction depth, $z$. We note that, in the few-electron regime of interest for these analyses, the S2 response may be regarded as a collection of single electron (SE) signals rather than the well-formed pulse shape observed at higher energies. A reasonable pulse width estimator is therefore the time separation between the first and the last SE signals in the S2 pulse.

The distribution of SE arrival times within the S2 pulse due to diffusion can be written as
\begin{equation}
    \label{eq:diffusion}
    P(t-t_d) = \frac{v_d}{\sqrt{4\pi D_L t_d}}e^{\frac{-v_d^2(t-t_d)^2}{4 D_L t_d}},
\end{equation}
where $t_d$ is the drift time, $v_d$ is the electron drift speed and the $D_L$ is the longitudinal diffusion coefficient for the liquid. Recent measurements~\cite{exodiffusion,HogenbirkDiffusion} obtained a diffusion constant $D_L \approx 25$~cm$^2$/s and electron drift speed $v_d \approx 1.7\textrm{~mm/$\mu$s}$ for the nominal LZ drift field, translating to $\sigma_z = \sqrt{2D_L\,t_{d}} = 2.1$~mm for electrons from the deepest events in LZ (equivalent to a 1.25~$\mu$s typical timing spread per ionization electron).

Convolving an idealized boxcar pulse shape applicable to the SE response with the Gaussian-distributed arrival times of individual electrons gives the probability distribution of individual photon detection times. The mean duration of the SE response is predicted to be 1.2~$\mu$s for nominal gate-anode conditions; this will vary with the radial position of the interaction due to electrostatic grid deflection, but this effect can be corrected for with calibration data. The SE timing response departs from the ideal boxcar function for several reasons. Firstly, the SE shape is influenced by local field enhancements due to grid geometry, as well as the possibility of some electrons overshooting the anode before returning (see Ch.~7 in Ref.~\cite{AdamThesis}). We do not account for these effects in this study. Furthermore, the width of SE pulses fluctuates due to the detection times of the individual photoelectrons within the pulse; this is further affected by the electronics response chain. Optical and electronics chain simulations confirm that maintaining a rectangular pulse shape is an adequate approximation for the SE response. This excludes diffusion, which applies only to multi-electron S2 pulses. We use a variable SE width with mean $\sigma_0=1.2~\mu$s and dispersion $\delta = 40~$ns.

The S2 timing spread is then calculated by sampling Eq.~\ref{eq:diffusion} for each electron in the S2 to simulate the effect of diffusion, and by adding the samples of the $\sigma_0$ distribution from the two electrons that picked up the most extreme diffusion spreads to simulate the effect of varying SE widths. While this is a conservative measure of the pulse width, it is sufficient to explore diffusion effects for S2 pulses composed of only a few electrons, for which more sophisticated methods do not introduce a significant improvement. Using this approach, cathode events at the 5-electron threshold show a mean S2 pulse width of $4.0~\mathrm\mu$s, with $90$\% of pulses in the range $2.3$--$5.7~\mathrm\mu$s (width distributions for cathode and other background sources are calculated and shown in Sec.~\ref{sec:S2oanalysis}).

A related consideration is the pulse shape for ionization emitted by the grids themselves. It has recently become apparent that the S2 pulse from these events is distorted compared to that of bulk events, and that these can be efficiently identified by dedicated analysis techniques~\cite{luxs2o}. We do not take advantage of this effect, which is likely to be helpful mostly for gate events, and instead assign to these events the same timing properties as those predicted by Eq.~\ref{eq:diffusion}.

Position reconstruction in the $x,y$ plane remains precise at the $5$-electron threshold. An LZ implementation of the Mercury algorithm~\cite{ZIIIMercury,LUXMercury} indicates that the smallest pulses in the S2-only analysis carry a $\sigma\!\approx\!1$~cm uncertainty in their radial position (improving to $\sim\!0.5$~cm for the larger signals in the 2-fold and nominal analyses). That introduces a modest $3\%$ ($\sim\!180$~kg) uncertainty in the fiducial mass considered in the S2-only case. This and other uncertainties, such as those associated with the transport of ionization in the detector and the generation of the S2 response, are expected to be subdominant and will not be considered here. We note, however, that most can be precisely studied in the low-energy regime through dispersed calibration sources, and especially with tritiated methane, which produces uniformly distributed events at low energies---above and below the S1 threshold---with a well-known energy spectrum~\cite{LUXCH3T}.

%%%%%%%%%%%%%%%%%%%%%%%%%%%%%%%%%%%%%%%%%%%%%%%%%%%%%%%%%%%%%%%%%%%%%%%%%%%%%
\section{Lowering the energy threshold of the LZ experiment}
In this section we apply the techniques introduced in Sec.~\ref{sec:techniques} to simulated data of the LZ experiment. In Sec.~\ref{sec:DPEanalysis} we use the DPE technique to lower the S1 coincidence requirement of the LZ analysis from 3 to 2-fold and calculate the sensitivity improvement. Analogously, in Sec.~\ref{sec:S2oanalysis} we drop the S1 coincidence requirement with the S2-only technique and present the gains in sensitivity that can be achieved.

\subsection{The DPE analysis \label{sec:DPEanalysis}}
To estimate the improvement in sensitivity of the experiment to DM-induced nuclear recoils using the DPE technique outlined above, we extended the signal and (relevant) background models in the standard simulation framework described in Ref.~\cite{LZsens} to smaller responses. For the signal model, the DM mass range explored is $\left[2.5, 12.0\right]$~GeV/c$^2$. For the background model, a realistic S2-only spectrum down to 5 extracted electrons is composed, using LUX data, for the S2-only analysis (described in Sec.~\ref{sec:S2oanalysis}), with 2-fold dark count coincidences faking S1 pulses randomly assigned to that spectrum for the DPE analysis. Electron lifetime corrections are applied to the S2 area distribution to calculate the PDF, which for random coincidence backgrounds are equivalent to a random-depth correction of the S2-only events.

\begin{figure*}[htbp]
  \centering
  \subfigure{\includegraphics[width=0.49\textwidth]{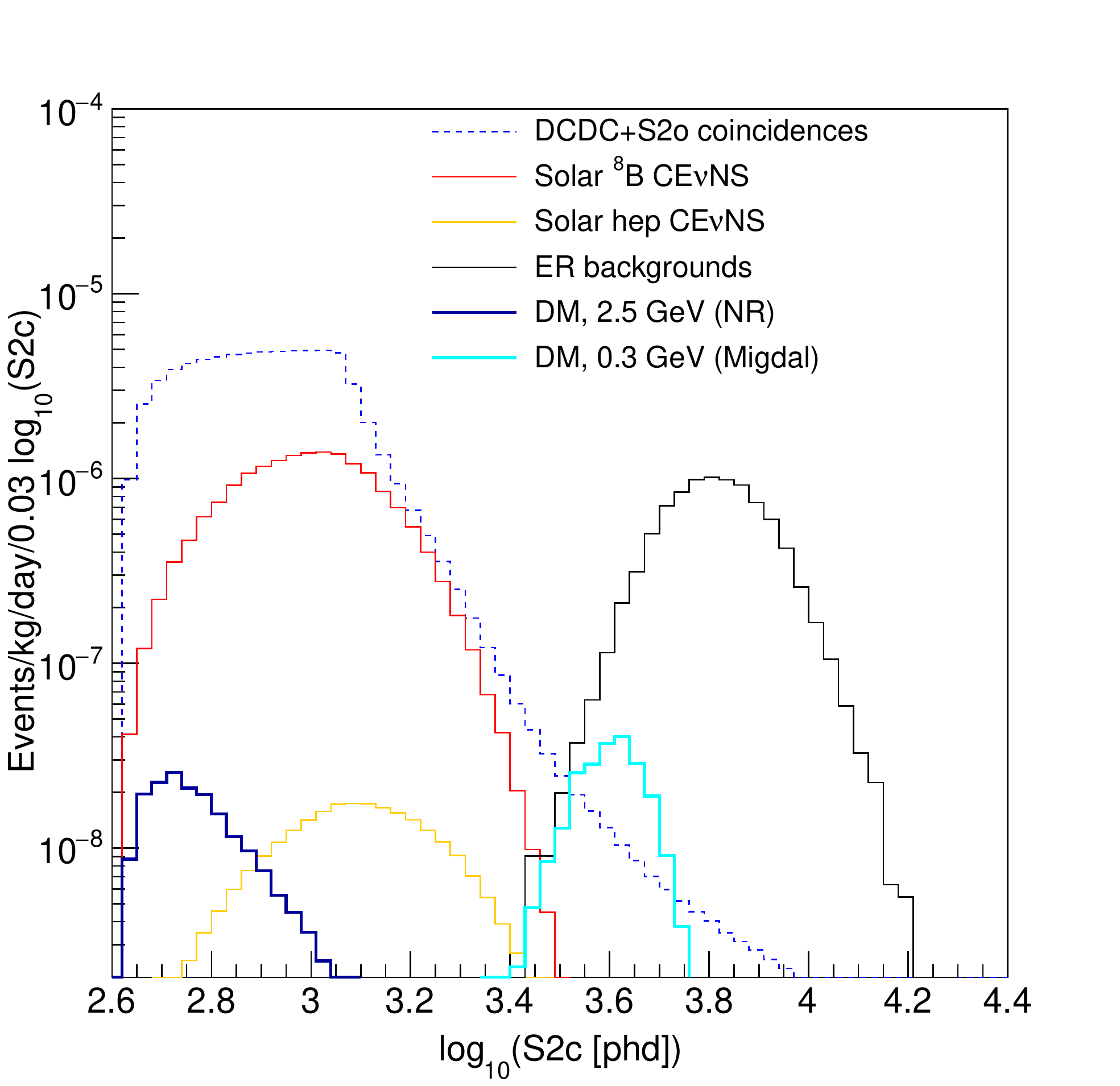}}\quad
  \subfigure{\includegraphics[width=0.49\textwidth]{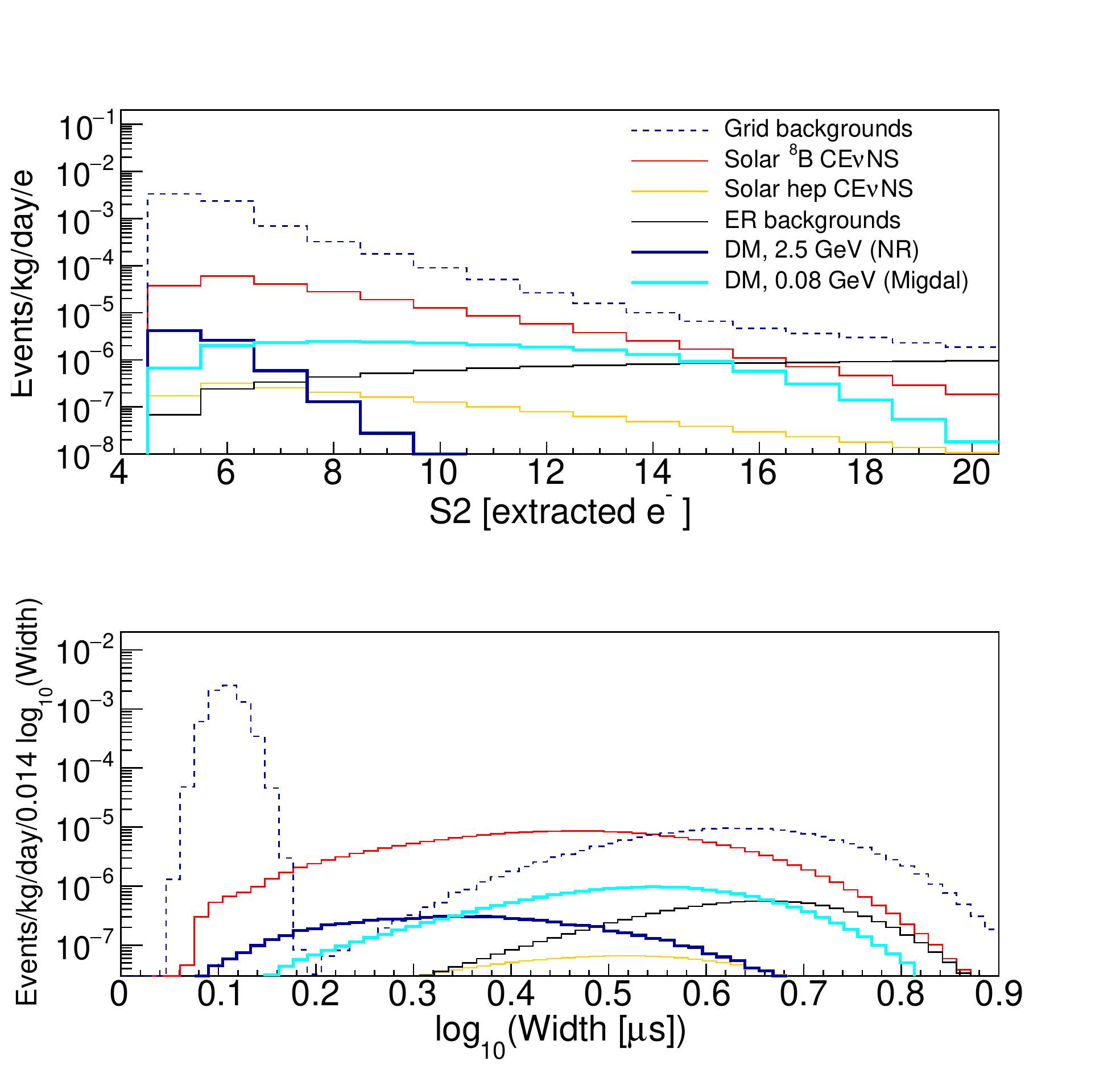}}
  \caption{Left: Spectra of the main components of the DPE analysis as a function of $\log_{10}(\textrm{S2c [phd]})$ for events with S1 signals with precisely two channels in coincidence. Only the $\textrm{S2r}>420$~phe cut has been applied. In the NR DM search, random coincidences of two dark counts and an isolated S2 pulse (labeled ``DCDC+S2o") dominate the background for small S2 areas--and low DM masses--but can be greatly reduced with the DPE and S1-window cuts (see Table~\ref{table:bkgcounts}). Solar $^{8}$B neutrino signals remain as an irreducible background for DM nuclear recoil searches, while $^{222}$Rn and other ER backgrounds dominate in the search for DM-induced electron recoils via the Migdal effect (see Section~\ref{subsec:DPEMigdal}). Right: Main components of the S2-only analysis projected onto the two variables used: S2 pulse size in extracted electrons (top) and S2 pulse width (bottom). Only the $\textrm{S2r}>420$~phe cut has been applied. The grid contributions dominate the background rate, but S2 shape differences can be exploited to recover sensitivity, as shown in the lower panel. The S2 width of background events originating at the gate tends to have smaller values than those originating at the cathode (see Fig.~\ref{fig:contourS2o}), and both distributions differ from that of interactions taking place over the whole analysis volume. An electron lifetime of $850~\mathrm{\mu}$s (one drift length) applies to all panels. \label{fig:SignBkgSpectra}}
\end{figure*}

In the nominal case ($\geq\!3$-fold S1) our likelihood-based statistical analysis relies on PDFs produced using NEST for both signal and background in S1--$\log_{10}$S2 space (here S1 and S2 represent pulse areas). Section~\ref{subsec:S1andDPE} explained how the PMT responses are modeled. For the special case of 2-fold signals passing the DPE cuts, no significant additional information is gained from the S1 area, and the PDFs are treated separately as unidimensional in $\log_{10}$S2---with the joint likelihood factorized into a DPE component and a $\geq\!3$-fold component (this is hereafter referred to as the `combined' analysis).

Figure~\ref{fig:SignBkgSpectra} (left) shows the PDFs for the DM signal and the main background components of the DPE analysis. These are plotted as a function of the corrected S2 area variable (S2c), the result of applying spatial corrections to the raw S2 area. This includes correcting for the $850~\mathrm{\mu}$s electron lifetime.

\begin{table}[tb]
\caption{Background counts in the standard 1,000 live-day exposure of a 5.6-tonne fiducial mass (unless indicated otherwise), with S2r$>$420~phe and different cuts applied to the S1 signal: for the S2-only analysis, for S1 pulses with exactly two coincident channels (`2-fold'), adding DPE cuts to the latter (`DPE'), and for the combination of nominal ($\geq3$-fold S1) and DPE analyses (`Comb'). The uncertainties associated with each source are shown on the rightmost column (`$\sigma/N$'). Individual background components are described in the text. ER PDFs are integrated over their full analysis range, shown in Fig.~\ref{fig:SignBkgSpectra}, and are indicative only.}
\label{table:bkgcounts}
\begin{tabular}{ c l r r r r r}
 \hline\hline
 \multicolumn{2}{c}{Background} &     \multicolumn{4}{c}{Counts after cuts} & \\
\hline
Type & Source & ~S2-only$^\dagger$ & ~~2-fold & ~~DPE & ~Comb$^{\dagger\dagger}$ & $\sigma/N$ \\
\hline
NR  & $\nu$~$^{8}$B\,$^\ddag$ & 1,390  & 100   &  15    & 42 & 4\%\\%&   27\\ 
    & $\nu$~hep     & 10     & 1.4  &  0.20  & 0.95 & 15\% \\%& 0.75\\
    & $\nu$~DSN     & 0.28   & 0.05 &  0.01  & 0.15 & 50\% \\%& 0.14\\
    & $\nu$~atm     & 0.22   & 0.05 &  0.01  & 0.64 & 25\% \\%& 0.63\\
    & materials     &    --  & 0.07 &  0.014 & 0.86 & 20\% \\%& 0.84\\ 
    \hline
ER  & $^{222}$Rn  &   44   &  32  & 4.6 & 1,790 & 10\% \\%& 1,790 \\
    & $^{220}$Rn  &   7.2  &  5.2  & 0.77 & 295 & 10\% \\%&   295\\
    & $\nu$~pp    &   12  &  9.3   & 1.3 & 572  & 2\% \\%&   571\\ 
    & $^{85}$Kr   &   1.7  &  1.4  & 0.30 & 78  & 20\% \\%&    78\\
    & $^{136}$Xe  &   0.03 &  1.2  & 0.18 & 428 & 50\% \\%&   428\\
    & materials   &     -- &  2.8  & 0.41 & 160 & 20\% \\%&   159\\
    \hline
Random & 2DC+S2o & 44,300 & 406 & 0.06 & 0.06 & 20\%\\%& -- \\
\hline\hline
\end{tabular}
\begin{flushleft}
    {\footnotesize 
    $^\ddag$ Counts for 1.7~ms lifetime: 2,020, 140, 21 and 58, respectively.\newline}
    $^\dagger$ 6.2-tonne analysis volume. See text for treatment of spatially non-uniform background components. \newline 
    $^{\dagger\dagger}$ Nominal contributions integrated over $\textrm{S1 [phd]}\leq100$ and $2.5\leq\log_{10}{\textrm{S2 [phd]}}\leq5.5$.
\end{flushleft}
\end{table}

In Table~\ref{table:bkgcounts} we list background counts for various S1 selection criteria, for the major background components, and explain below how these are determined. The backgrounds applicable to this analysis are treated as in Ref.~\cite{LZsens}. Spatially uniform backgrounds include the coherent nuclear scattering of solar ($^8$B, hep), atmospheric and diffuse supernova neutrinos (NR) and elastic $\nu$-$e$ scattering of solar pp neutrinos, `naked' betas from radon progeny at the nominal 2~$\mu$Bq/kg total specific activity, $^{85}$Kr beta decays at 0.015~ppt~(g/g) trace concentration, and the $2\nu2\beta$ decay of $^{136}$Xe~\cite{ExoXe136,KamlandXe136} (ER). Non-uniform backgrounds from detector materials and surfaces are also treated as in Ref.~\cite{LZsens}, since the DPE analysis preserves the fiducialization power of the nominal S1-S2 analysis.

In the radial direction we consider the same fiducial volume definition as in the nominal analysis, cutting at 40~mm from the TPC walls. For events with an S1 pulse the mis-reconstruction of wall interactions into this fiducial volume is expected to be negligible at the S2 threshold (cf.~Section~3.6.5 in Ref.~\cite{LZTDR}).

\begin{figure*}[htp]
  \centering
  \subfigure{\includegraphics[width=0.49\textwidth]{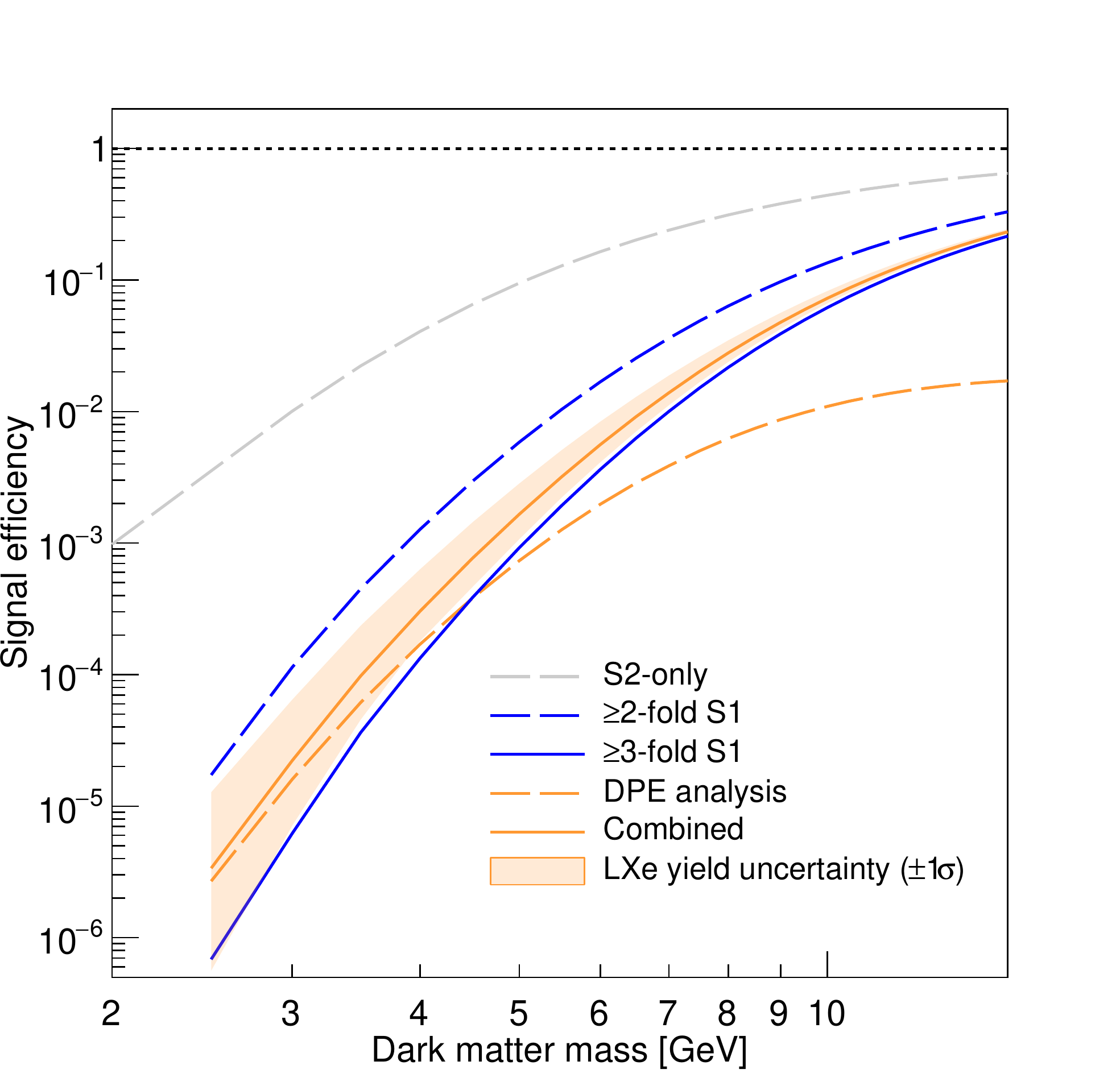}}\quad
  \subfigure{\includegraphics[width=0.49\textwidth]{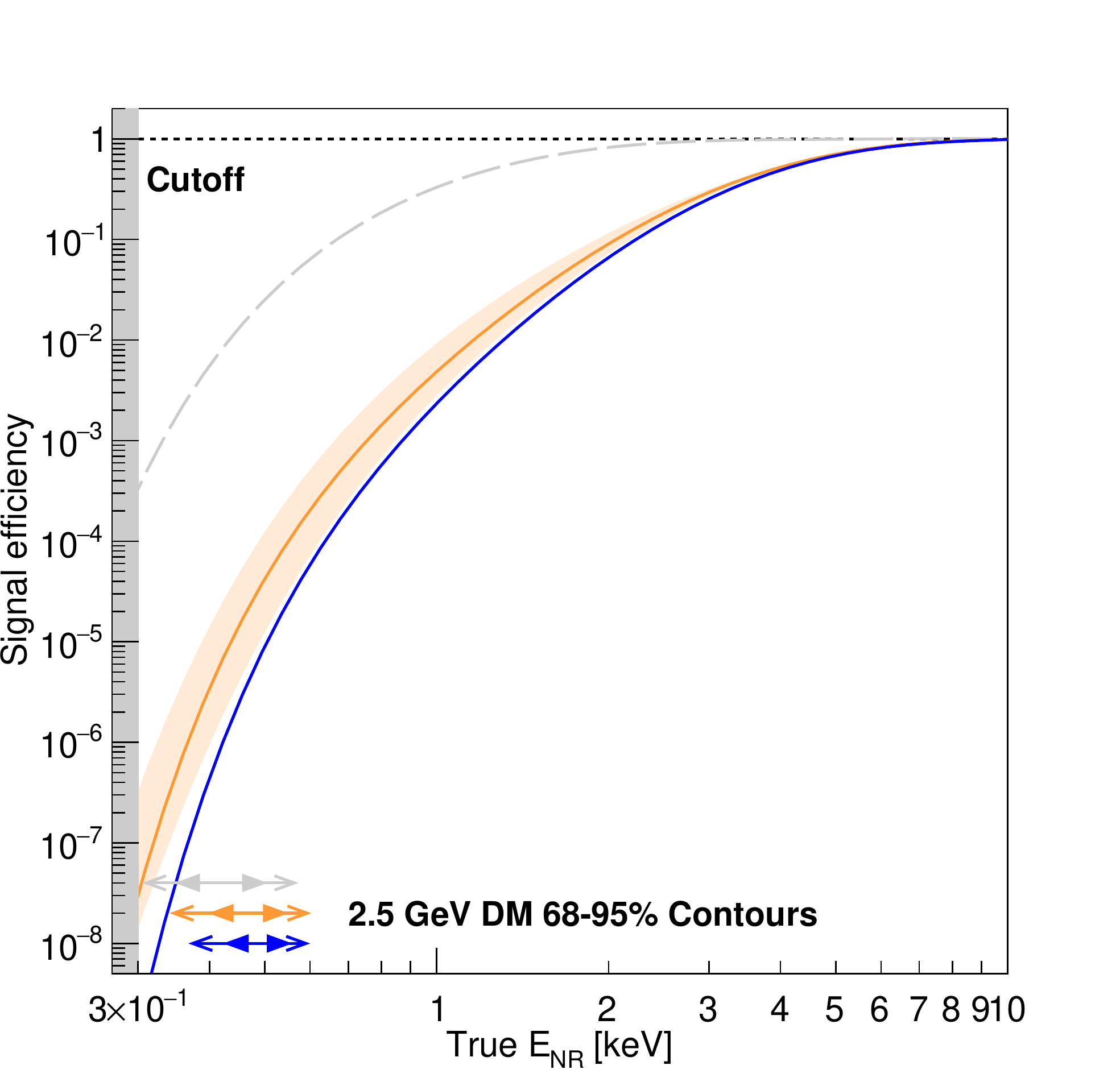}}
  \caption{Left: Dark matter signal efficiency as a function of particle mass for different cuts. The S2-only analysis (dashed gray) requires only S2r$>\!\! 420$~phe. The $\geq\!2$-fold S1 selection (dashed blue) is shown for comparison with the nominal $\!\geq\!3$-fold S1 analysis selection (solid blue). The DPE analysis (dashed orange) includes events with an S1 coincidence of precisely 2 channels, passing the DPE and S1-window cuts. Below 5~$\textrm{GeV/c$^2$}$, the DPE analysis efficiency is higher than that at $\geq\!3$-fold even after the DPE and S1-window cuts. The combined efficiency of DPE and nominal analyses is also shown (solid orange), with the shaded area representing the $1\sigma$ LXe yield uncertainty. Right: Efficiency as a function of true recoil energy for the S2-only, combined and nominal analysis (same color scheme). The closed and open arrows indicate the $68\%$ and $95\%$ contours of the effective recoil energy distribution of a 2.5~GeV/c$^2$ dark matter particle for their respective analysis (same color and relative positioning). The model cutoff at 0.3~keV is shown as a gray band. \label{fig:effSignal} }
\end{figure*}

The signal model assumes the standard dark matter halo with the same parameters used in Ref.~\cite{LZsens}. The signal efficiency is shown in Fig.~\ref{fig:effSignal} as a function of DM particle mass $m_{\rm DM}$ (left) and true recoil energy (right). The DPE analysis efficiency overtakes the $\geq\!3$-fold signal efficiency for $m_{\rm DM}\leq\!5$~GeV/c$^2$ even after all analysis cuts. At $m_{\rm DM}=2.5$~GeV/c$^2$, it becomes 4 times larger than the nominal efficiency, which reaches values as low as $7 \times 10^{-7}$. The recoil energy cutoff at $0.3$~keV does not affect the models considered in the analysis, but clearly new measurements are desirable to validate the light yield predictions of the NEST model, currently measured for nuclear recoil energies only as low as $1.08$~keV---but with new results confirming the yields at 0.45~keVnr~\cite{Huang:2020ryt}.

Table~\ref{table:bkgcounts} highlights also how the number of $^8$B counts evolves with progressive S1 selection cuts. At nominal electron lifetime, the number of detected events in the nominal analysis is $\approx\!27$, increasing by $15$ counts by adding the DPE population (out of a total of $100$~counts with 2-fold coincidences).

\subsubsection{Nuclear recoil search sensitivity}\label{subsec:DPENR}

Profile likelihood ratio (PLR) tests are performed using the PDFs and efficiencies described above. We treat background rate uncertainties as nuisance parameters in the likelihood function. We follow Table~IV in Ref.~\cite{LZsens} for our background rates, and we add a random coincidence background, as described in the previous section.

\begin{figure}[ht]
\includegraphics[width=0.5\textwidth]{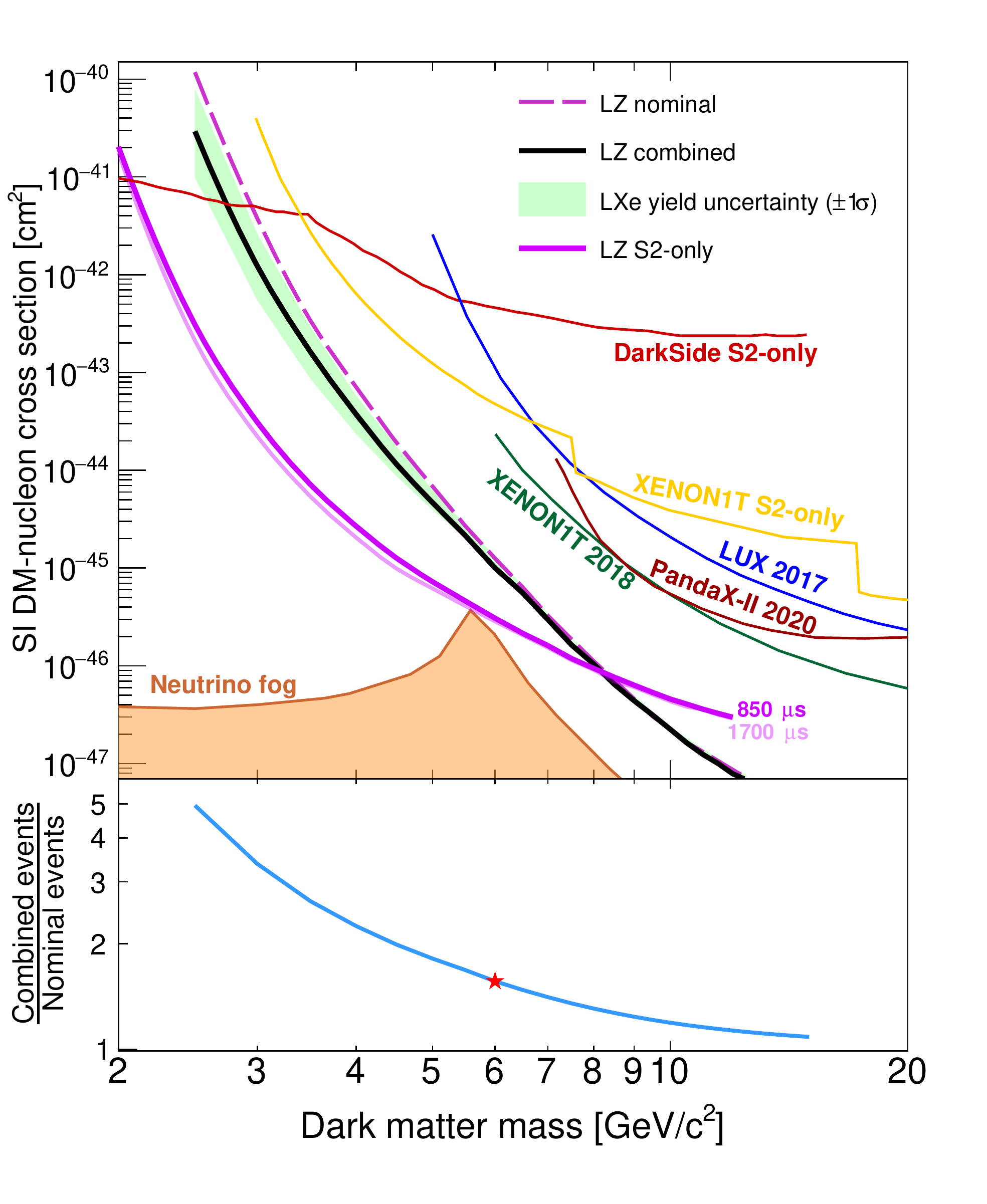}
\caption{Upper panel: Spin-independent sensitivity for the nominal analysis~\cite{LZsens} (violet, dashed), the DPE plus nominal combination analysis (black, solid), and for the S2-only analysis (violet, solid). The neutrino fog is shown for an idealized xenon detector~\cite{NeutrinoFloor1, NeutrinoFloor2}. Also shown are experimental limits published by DarkSide-50 S2-only (red line)~\cite{darkside50}, LUX 2017 (blue, solid)~\cite{luxRun3p4} and XENON1T (green, solid)~\cite{Xenon1T}.
The uncertainty introduced by the LXe yields is shown as a green band around the combined result.
The effect of doubling the electron lifetime is shown for the S2-only analysis (dimmed violet). Lower panel: ratio of signal efficiency between the combined and nominal LZ analyses (blue, solid) with the ratio corresponding to the solar $^8$B neutrino rate highlighted (red star).\label{fig:WIMPsens}}
\end{figure}

The combination (product) of DPE and $\geq\!3$-fold likelihoods is used to derive the spin-independent sensitivity shown in Fig.~\ref{fig:WIMPsens}.
The sensitivity for the combined analysis represents a factor 4 improvement at 2.5~GeV/c$^2$ with respect to the nominal analysis. This improvement in signal efficiency with the inclusion of DPE signals is illustrated on the lower panel with the $^8$B neutrino rate (indicated by a red star) increasing by $60\%$. This improved efficiency opens up the possibility of a more precise $^8$B neutrino detection via CE$\nu$NS. However, the overlap of the $^8$B
 spectrum with the DM signal spectra also limits severely the improvement in DM sensitivity attainable in this mass region.
 
 A number of systematic effects were analyzed separately, rather than incorporated as nuisance parameters in the PLR framework. A small contribution to the overall uncertainty is obtained from the measurement of the electron lifetime. We consider a $5\%$ uncertainty in this measurement, informed by the experience of XENON1T~\cite{Xe10S2o}. Although LUX achieved $<\!1$\%~error in each measurement of the lifetime (cf.~Fig.~23 in Ref.~\cite{LUXPRD}), changes in xenon purity may occur between measurements in a long dataset. The effect of this uncertainty was tested in the PLR analysis and found to be negligible.
 
 For very low masses, uncertainties in the LXe scintillation and ionization yields significantly affect the sensitivity of the experiment. The NEST model was varied within its expected uncertainties to account for this effect, as discussed in Sec.~\ref{subsec:S1andDPE}. We note that the largest uncertainties are associated with the scintillation yield, while the ionization yield is well measured above $0.3$~keV. Scintillation yield variations move events between the $\leq\!1$-fold, DPE and $\geq\!3$-fold categories. Hence, we have considered this uncertainty by recalculating the combined model sensitivity for a $\pm1\sigma$ variation of the yields. This uncertainty is represented by the green band in Fig.~\ref{fig:WIMPsens}.
 
 Another significant parameter is the rate of accidental coincidences. We find that similar signal-to-background levels are attained with small adjustments of the DPE cuts. In particular, we fully recalculated the sensitivity with a $5\%$ lower DPE area cut and no difference was found. Therefore we do not consider variations of the S2-only and PMT dark count rates.
 
 Finally, we repeated the analysis for the higher electron lifetime scenario (1.7~ms), this being a key parameter when dealing with ionization signals at threshold. The rate of low-energy NR signals uniformly distributed in the active volume increases noticeably (for example, the $^8$B rate increases by $40\%$ as indicated in the footnote to Table~\ref{table:bkgcounts}, but the ratio of $^8$B events between the nominal and combined DPE analyses remains constant, since both rates grow equally). This affects the dominant background from cathode events differently. With improved purity fewer electrons are lost, and the expected event rate grows by a factor of 5. The lifetime correction applied to each event is smaller on average, producing a softer spectrum than at $850$~$\mu$s. Overall, however, the analysis is unaffected by random coincidence backgrounds after DPE cuts. The cross-section sensitivity improves roughly with $s/\sqrt{b}$, giving a factor $\sim1.6$ improvement for $m_{\rm DM} = 2.5$~GeV/c$^2$.
 
 \subsubsection{Electron recoil searches for sub-GeV particles}\label{subsec:DPEMigdal}
 
\begin{figure}[ht]
\includegraphics[width=0.5\textwidth]{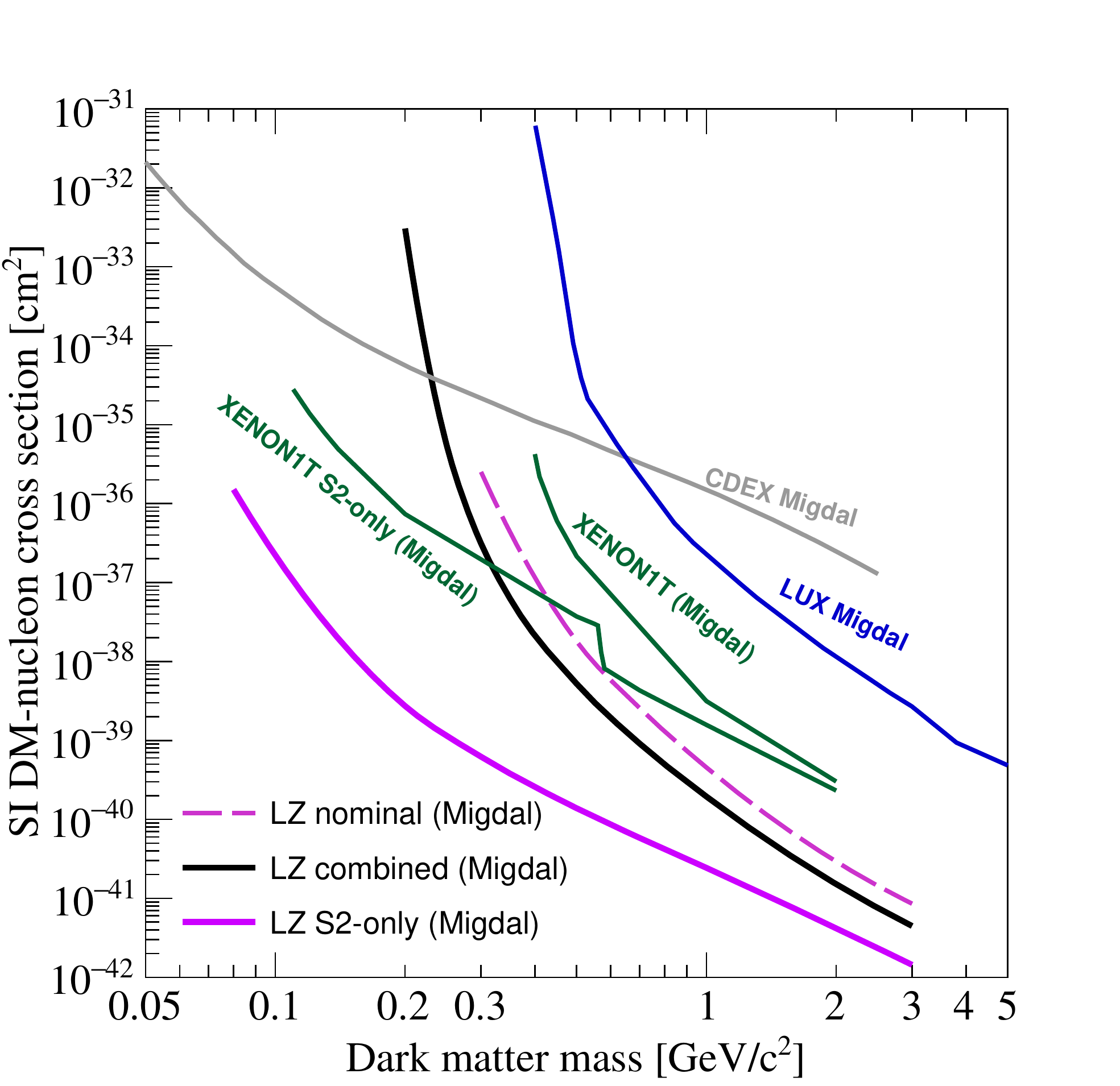}
\caption{Spin-independent sensitivity for ER-inducing Migdal effect (heavy scalar mediator case). The nominal LZ sensitivity at $\geq\!3$-fold (violet, dashed), the sensitivity of the DPE analysis (black, solid) and the S2-only sensitivity (violet, solid) are shown. An order of magnitude improvement is obtained at $0.3$~GeV/c$^2$ DM mass in the DPE case, while the S2-only analysis allows to extend the search below $100$~MeV/c$^2$. Experimental limits from similar analyses in LUX (blue, solid)~\cite{LUXMigdal}, XENON1T~\cite{Xe1TMigdal} and CDEX (gray, solid)~\cite{CDEXMigdal} are also shown.\label{fig:Migdalsens}}
\end{figure}

 The detection of dark matter particles recoiling against the relatively heavy xenon nucleus is presently limited to dark matter masses $m_{\rm DM}\gtrsim2$~GeV/c$^2$. These are the lightest particles trapped in the galactic halo that can produce a xenon recoil of $0.3$~keV. It has been noted, however, that there is a small probability for the recoiling nucleus to emit a bremsstrahlung photon or even to induce a change in the atomic energy levels and force the emission of an atomic electron via the so-called ``Migdal effect"~\cite{MigdalEffect,Dolan2018}. In $\lesssim 10^{-5}$ of the collisions, a sub-GeV DM particle may produce an undetectable nuclear recoil accompanied by a larger electron signal above the ER threshold. It should be remarked that experimental confirmation of the Migdal effect in any medium is still lacking.
 
 Using a similar approach to that adopted by LUX~\cite{LUXMigdal}, we perform a sensitivity study using baseline LZ parameters for the detection of light dark matter particles in the mass range $0.3\!-\!5.0$~GeV/c$^2$ depositing energy in the LXe target via the emission of a Migdal electron. We simulate the energy deposition in the detector following the spectra from Ref.~\cite{MigdalEffect}. The fraction of energy observed via the initial nuclear recoil is smaller than the yield fluctuations of the ER recoil signal, and hence it is neglected in this study. Apart from the usual heavy scalar mediator case assumed in standard dark matter searches, Migdal searches have explored other types of mediators. We show results for the heavy scalar mediator case for illustration, but the analysis is equivalent for other types of mediators. The threshold of the nominal ($\geq\!3$-fold) analysis is extended as previously.

Following a cut optimization procedure analogous to that described in Sec.~\ref{subsec:S1andDPE} for the DPE analysis, we find that the ER backgrounds dominate over random coincidences of dark counts and S2-only events in this case and no additional cuts are needed. This case is also shown in Fig.~\ref{fig:SignBkgSpectra}. In practice, additional cuts may be implemented should a higher rate of random backgrounds be found. The result of the analysis is shown in Fig.~\ref{fig:Migdalsens}.

The reach of LZ can be extended down to sub-GeV DM masses by exploiting the Migdal effect. The relaxation of the straight 3-fold S1 requirement to 2-fold with DPE improves the sensitivity by about an order of magnitude but lowering the low-mass threshold even further is limited by the ER light yield.

\subsection{The S2-only analysis \label{sec:S2oanalysis}}

The sensitivity achieved with an S2-only analysis is estimated in the mass range $\left[2.0, 12.0\right]$~GeV/c$^2$. The signal and background models are extended to include events irrespective of their S1 coincidence level. The S2 pulse width is used to estimate the electron drift time; this parameter is simulated as described in Sec.~\ref{subsec:S2only}. PDFs are produced in the S2--$\log_{10}$(width) space. The smallest S2 signals contribute the largest gains in sensitivity, hence only S2 pulses up to 20 extracted electrons are considered.

\begin{figure}
\includegraphics[width=0.5\textwidth]{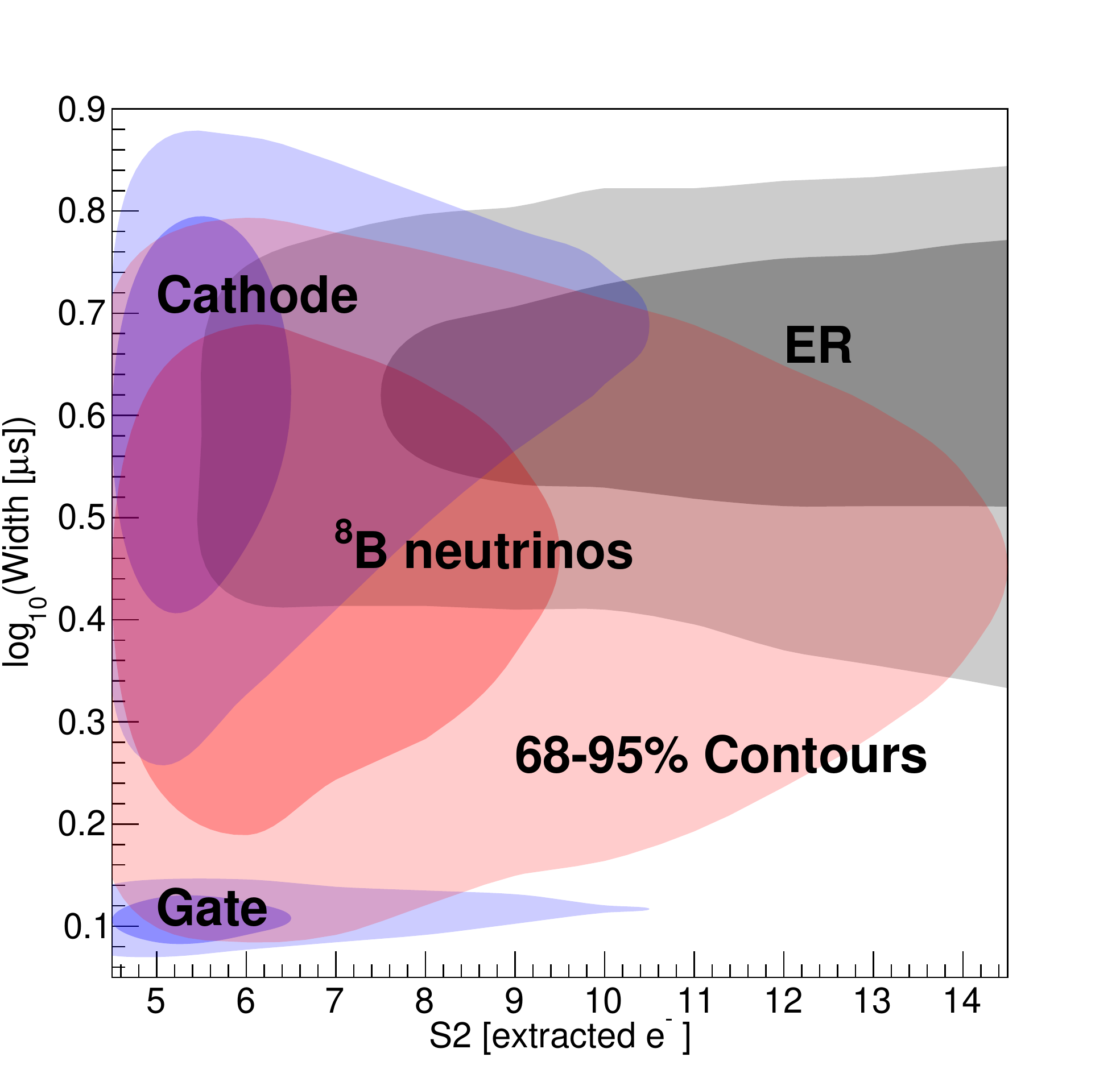}
\caption{Two-dimensional contours (68\%, 95\%) of selected PDFs from the S2-only analysis. The $^8$B-neutrino distribution (and all other NR signals) overlaps the cathode, gate and ER distributions, but it crucially dominates at intermediate values of the S2 width due to the difference in its spatial distribution and energy spectrum. \label{fig:contourS2o}}
\end{figure}

Figure~\ref{fig:SignBkgSpectra} (right) shows the PDFs for the DM signal and the main background components of the S2-only analysis projected onto the S2 pulse area and S2 pulse width axes, while Fig.~\ref{fig:contourS2o} shows two-dimensional 68\% and 95\% contours for relevant background components of the analysis. In this instance the radial position can still be reconstructed (and used for S2 corrections), but the depth coordinate cannot be precisely determined in the absence of an S1 pulse. In this case, a hard fiducialization cut in the $z$ coordinate from the S2 pulse width alone would be too punitive. For this reason, we do not apply an S2-width cut and instead use it as an input into the PLR, which increases the analysis mass from $5.6$~tonnes to $6.2$~tonnes (the full vertical extent of the detector).
Clearly, the cathode background events are especially challenging to discriminate due to maximal diffusion smearing, as the S2 pulse width distribution in Fig.~\ref{fig:SignBkgSpectra} clearly highlights. However, the mean of the width distribution for a given number of S2 electrons is higher for cathode events than for bulk events, allowing for some discrimination, especially for the larger S2 pulses, as shown in Fig.~\ref{fig:contourS2o}.

The S2-only background counts are listed in Table~\ref{table:bkgcounts}. The spatially uniform backgrounds are treated as in the DPE analysis. In contrast, the treatment of non-uniform backgrounds from detector materials and surfaces is different in the S2-only analysis. The model considered in Ref.~\cite{LZsens} would not be representative in the larger analysis volume specifically considered in the S2-only study. Grid backgrounds, rather than external radiation, heavily dominate the background model once $z$ fiducialization is lifted, and so the precise spatial distribution of the radiological backgrounds is of subdominant importance.

The S2-only background spectrum used in both the DPE and S2-only analyses is derived from that observed in LUX~\cite{LUXDPE} at high electron lifetime, scaled by the ratio of the S2 detection efficiencies of the two detectors, and assigned equally to the gate and cathode grids. The LZ cathode spectrum is then corrected for electron lifetime (spectral distribution and normalization) under the two xenon purity scenarios considered in this study. Only $\sim3\%$ of cathode events pass all analysis cuts for a $850$~$\mu$s lifetime, increasing to $\sim16\%$ for $1.7$~ms.

Radial fiducialization is also implemented as in the nominal analysis, cutting $40$~mm from the TPC walls. This will be more challenging in the S2-only analysis, as the stray field components cannot be mapped accurately as a function of depth, and this will require robust calibration.

In Table~\ref{table:bkgcounts} the $^8$B counts for the S2-only analysis are shown and compared to the other S1-selection criteria. For the larger 6.2-tonne analysis volume this is $1,390$ counts, two orders of magnitude larger than in the nominal and DPE analyses.

\subsubsection{Nuclear recoil search sensitivity}\label{subsec:S2oNR}
A PLR test is performed using the PDFs and efficiencies just described. We treat the uniform PDFs in an equivalent way to the DPE analysis. For the particular case of the S2-only background rate, we assume that it can be measured with $20\%$ uncertainty. Other systematic uncertainties are discussed below.

The study is performed on DM masses down to $2.0$~GeV/c$^2$. The spin-independent sensitivity is shown in Fig.~\ref{fig:WIMPsens}. It represents a factor $\sim200$ improvement with respect to the nominal analysis at $2.5$~GeV/c$^2$.

We recognize that previous liquid xenon experiments have found challenging to model S2-only backgrounds rigorously, and have therefore set upper limits based on single-sided statistical techniques~\cite{Xe10S2o, Xe1TS2o}. Therefore, it is essential to achieve a model of sufficient precision if LZ is to have discovery power through such an analysis. With reference to Fig.~\ref{fig:contourS2o}, it is clear that cathode events are the main concern, given their broad S2-width distribution (in contrast, the gate emissions are easier to reject from S2 pulse width). Rejecting cathode events should be easier in a longer detector. In addition, cathode events may be studied with a 1-fold coincidence requirement in S1 once the precise drift time from the grid position is determined from larger signals. Two more avenues may be explored to help determine the cathode event spectrum: the dependencies on electron lifetime and on drift field, which are distinct from those for bulk events. Finally, unblinding a small fraction ($\sim10\%$) of the data may be considered to fit the cathode spectrum.

We analyzed the same systematic uncertainties outside of the PLR framework that we considered for the DPE analysis. 
The uncertainty induced by the electron lifetime measurement is also found to be negligible. 
The LXe scintillation and ionization yield uncertainties are higher at the masses most relevant to the S2-only analysis. However, since the most significant uncertainty is associated with the scintillation yield, the impact on the S2-only analysis is much smaller than on the nominal analysis.

We also repeat the analysis assuming a higher electron lifetime ($1.7$~ms). The improvement observed is a factor of $\lesssim1.3$ with respect to $850$~$\mu$s, hindered by the increased survival probability of cathode events.

Finally, the uncertainty on the mean S2-only background rate is explored by increasing the nominal $1$~mHz rate by a factor of 10. While this is a significant uncertainty, its effect is much reduced due to the discrimination between bulk and cathode events gained from the S2 width. As shown in Fig.~\ref{fig:SignBkgSpectra} (lower right panel), $^8$B dominates in the region where the DM spectrum peaks. The effect of this systematic uncertainty becomes significant at $2$~GeV/c$^2$ (a factor of $\sim0.9$ on the sensitivity), but is much smaller for higher masses.

\subsubsection{Electron recoil searches for sub-GeV particles}\label{subsec:S2oMigdal}

Extending the Migdal analysis to lower masses is possible when relaxing the S1 requirement completely. The S2-only analysis can bring about gains in sensitivity of several orders of magnitude at low masses due to the increased signal efficiency, previously limited by the LXe scintillation light yield. We perform this analysis with nominal grid emission and electron lifetime for dark matter masses $m_{\rm DM}\geq80$~MeV/c$^2$.

The results, shown in Fig.~\ref{fig:Migdalsens}, show gains of at least 2 orders of magnitude in the spin-independent cross-section sensitivity below $m_{\rm DM} = 0.3$~GeV/c$^2$ with respect to the current best limit~\cite{Xe1TMigdal}.

%%%%%%%%%%%%%%%%%%%%%%%%%%%%%%%%%%%%%%%%%%%%%%%%%%%%%%%%%%%%%%%%%%%%%%%%%%%%%
\section{Conclusions\label{sec:conclusions}}

The energy threshold considered when deriving the predicted spin-independent dark matter sensitivity of LZ has been lowered by relaxing the scintillation threshold of the analysis. Two cases were explored: a reduction from a 3- to 2-fold coincidence requirement in those cases where double photoelectron emission takes place in a photomultiplier, and an S2-only analysis which drops the S1 requirement altogether.

To account for the increased background due to random coincidences of PMT dark counts with S2-only events in the DPE analysis, timing and area cuts on the 2-fold S1 pulse have been optimized. The results lead to a factor $4$ improvement in sensitivity at $2.5$~GeV/c$^2$. The improvement at low masses is limited by the presence of $^8$B neutrinos and it is approximately equal to the change in $s/\sqrt{b}$. However, as shown in Figs.~\ref{fig:effSignal} and~\ref{fig:WIMPsens} (lower panel), the higher efficiency allows exploitation of the tail of the efficiency onset with lower uncertainty to observe nearly double the number of $^8$B neutrinos interacting via CE$\nu$NS---improving the discovery prospects.

The S2-only analysis achieves significant rejection of grid backgrounds by using S2 pulse width information, improving the LZ sensitivity at $2.5$~GeV/c$^2$ by $2$ orders of magnitude with respect to the combined analysis, and extending the reach to lower masses. This also gives a similar increase in the measured numbers of $^8$B interactions. The analysis is relatively immune to the rate of grid events and the electron lifetime. LZ is sufficiently large a detector such that the PLR analysis is able to distinguish these distributions from those arising in the LXe bulk.

The sensitivity to DM particles inducing the atomic Migdal effect has been explored as a way to extend the experiment's reach to sub-GeV masses further. Similar analyses were conducted with the DPE and S2-only techniques, lowering the dark matter mass threshold to $80$~MeV/c$^2$ with the latter due to the increased signal efficiency, and improving the sensitivity by up to three orders of magnitude for the masses within reach of the nominal analysis, down to $0.3$~GeV/c$^2$.

The DPE and S2-only analyses require a good understanding of different aspects of the detector. While the DPE analysis relies on the S2-only rate to tune the cuts, factors of $2$ uncertainty in that rate do not change the values significantly. This technique therefore represents a straightforward extension of the nominal analysis with modest gains in sensitivity, provided that the PMT response model, and in particular the DPE probability, is well known. Even lower S2-only and/or PMT dark rates would also allow the analysis to rely only on an S1 window cut provided that the timing calibration is precisely understood. In contrast, the S2-only analysis relies on an accurate characterization of background sources producing S2-only signals, which are the main contribution to the systematic uncertainty in the sensitivity. Finally, we highlighted the role played by achieving (and measuring accurately) the electron lifetime during the experiment, which will require frequent calibrations and interpolation between data points.

In conclusion, we have demonstrated the potential of LZ to operate at a lower scintillation threshold using two complementary approaches. This potential is shared by current and future two-phase xenon experiments and relies on the detailed understanding of the S2-only backgrounds and the PMT response model.

\begin{acknowledgments}
The research supporting this work took place in whole or in part at the Sanford Underground Research Facility (SURF) in Lead, South Dakota. Funding for this work is supported by the U.S. Department of Energy, Office of Science, Office of High Energy Physics under Contract Numbers DE-AC02-05CH11231, DE-SC0020216, DE-SC0012704, DE-SC0010010, DE-AC02-07CH11359, DE-SC0012161, DE-SC0014223, DE-SC0010813, DE-SC0009999, DE-NA0003180, DE-SC0011702,  DESC0010072, DE-SC0015708, DE-SC0006605, DE-SC0008475, DE-FG02-10ER46709, UW PRJ82AJ, DE-SC0013542, DE-AC02-76SF00515, DE-SC0018982, DE-SC0019066, DE-SC0015535, DE-AC52-07NA27344, \& DOE-SC0012447.    This research was also supported by U.S. National Science Foundation (NSF); the U.K. Science \& Technology Facilities Council under award numbers, ST/M003655/1, ST/M003981/1, ST/M003744/1, ST/M003639/1, ST/M003604/1, ST/R003181/1, ST/M003469/1, ST/S000739/1, ST/S000666/1, ST/S000828/1, ST/S000879/1, ST/S000933/1, ST/S000747/1, ST/S000801/1 and ST/R003181/1 (JD); Portuguese Foundation for Science and Technology (FCT) under award numbers PTDC/FIS-­PAR/28567/2017; the Institute for Basic Science, Korea (budget numbers IBS-R016-D1). We acknowledge additional support from the STFC Boulby Underground Laboratory in the U.K., the GridPP~\cite{gridpp1, gridpp2} and IRIS Collaborations, in particular at Imperial College London and additional support by the University College London (UCL) Cosmoparticle Initiative. This research used resources of the National Energy Research Scientific Computing Center, a DOE Office of Science User Facility supported by the Office of Science of the U.S. Department of Energy under Contract No. DE-AC02-05CH11231. The University of Edinburgh is a charitable body, registered in Scotland, with the registration number SC005336. The assistance of SURF and its personnel in providing physical access and general logistical and technical support is acknowledged.

\end{acknowledgments}

\bibliography{main}

%%%%%%%%%%%%%%%%%%%%%%%%%%%%%%%%%%%%%%%%%%%%%%%%%%%%%%%%%%%%%%%%%%%%%%%%
%END OF READABLE TEXT%%%%%%%%%%%%%%%%%%%%%%%%%%%%%%%%%%%%%%%%%%%%%%%%%%%
%%%%%%%%%%%%%%%%%%%%%%%%%%%%%%%%%%%%%%%%%%%%%%%%%%%%%%%%%%%%%%%%%%%%%%%%

\end{document}